# Hiding in the crowd: Spectral signatures of overcoordinated hydrogen bond environments

Tobias Morawietz[a], Andres S. Urbina[b], Patrick K. Wise[b], Xiangen Wu[c], Wanjun Lu[c], Dor Ben-Amotz[b], and Thomas E. Markland[a*]

a) Department of Chemistry, Stanford University, Stanford, CA 94305, United States

b) Department of Chemistry, Purdue University, West Lafayette, IN 47907, United States

c) College of Marine Science and Technology, China University of Geosciences, Wuhan 430074, China

*Email: tmarkland@stanford.edu

**Abstract**

Molecules with an excess number of hydrogen-bonding partners play a crucial role in fundamental chemical processes, ranging from the anomalous diffusion in supercooled water to the transport of aqueous proton defects and the ordering of water around hydrophobic solutes. Here we show that overcoordinated hydrogen bond environments can be identified in both the ambient and supercooled regimes of liquid water by combining experimental Raman multivariate curve resolution measurements and machine learning accelerated quantum simulations. In particular, we find that OH groups appearing in spectral regions usually associated with non-hydrogen-bonded species actually correspond to hydrogen bonds formed in overcoordinated environments. We further show that only these species exhibit a turnover in population as a function of temperature, which is robust and persists under both constant pressure and density conditions. This work thus provides a new tool to identify, interpret, and elucidate the spectral signatures of crowded hydrogen bond networks.

Overcoordinated hydrogen bond environments have been implicated as fundamental structures in the chemical dynamics of processes ranging from the anomalous diffusion in supercooled water,[1] to the reorientation mechanism of liquid water,[2] the immobilization of water molecules around hydrophobic solutes[3] and the transport of proton defects in aqueous systems.[4] In liquid water they occur when a water molecule that already possesses two donor and two acceptor hydrogen bonds receives a further hydrogen bond, transforming it to a double-donor triple-acceptor species. Although the importance of these hydrogen bonds has been recognized, their spectral signatures and structural properties, as well as their occurrence over a wide range of temperatures, have not previously been determined. Vibrational spectroscopy,[5,6] provides an opportunity to address these questions, given its sensitivity to variations in hydrogen-bond strength[7] and tetrahedrality,[8] as well as non-hydrogen bonded OH groups. However, decomposing the associated spectral signatures within the broad OH stretching region remains an outstanding challenge. Commonly employed deconvolution techniques assume a predefined shape of the sub-bands and depend on their assumed number, position and/or width.[9–11] A decomposition method that avoids such arbitrary and unphysical assumptions, and hence alleviates the associated pitfalls, is self-modeling curve resolution (SMCR), which provides a means of decomposing a collection of spectra into spectral components with exclusively positive intensity whose shape is determined entirely by the experimental spectra.[8,12–15]

Here, we demonstrate that SMCR decompositions of the Raman spectrum of water in the ambient and supercooled regimes reveal that, in addition to previously identified spectral features whose populations increase monotonically with decreasing temperature,[8,16] one can also identify features that exhibit a more interesting non-monotonic temperature dependence with a maximum at a temperature around 280 K. In particular, these spectral features include a previously unrecognized distinct peak in the high frequency part of the OH stretching region (at ~3640 cm$^{-1}$) which is normally associated with OH groups not involved in a hydrogen bond (so called "free" or "dangling" OH's). Using simulations based on machine-learned potentials[8,17–20] trained to water's electronic structure we demonstrate that this region of the decomposed spectrum corresponds to OH groups participating in overcoordinated hydrogen bonding environments. Even though the molecules in these overcoordinated environments satisfy the geometric criteria commonly associated with the formation of a hydrogen bond their OH vibrational frequencies are almost identical to those of non-hydrogen bonded (free) OH groups. By analyzing the correlations in the full vibrational spectrum obtained from the simulations we further predict the corresponding librational region in which these overcoordinated species can be observed and show that the decomposed anisotropic Raman spectrum allows these spectral signatures to be resolved experimentally. This analysis, which combines experimental and simulated temperature dependent Raman spectra of water in the liquid and supercooled regimes delivers the first unambiguous assignment of spectral signatures, in both the low- and high-frequency region, arising from overcoordinated hydrogen bond environments, thus providing a means of identifying them in other systems of chemical and biological relevance.

Figure 1a shows the OH stretch region of the experimental temperature-dependent isotropic Raman spectrum of liquid water, measured at temperatures from the supercooled regime to its boiling point (238 to 370 K). The temperature variation of the OH stretch region is captured by our simulations (Figure 1b) employing a machine-learned potential[8] whose results closely resemble the experimental results, although the predicted peaks are all shifted to higher frequencies by ~100 $cm^{-1}$. To investigate the origin of the temperature dependence of the OH stretching region we employ SMCR decomposition of the spectra into multiple spectral components, beginning with a global two-component decomposition revealing monotonically temperature dependent features, followed by the more detailed multi-component analysis that reveals the third non-monotonic feature whose existence and structural assignments is primary finding and focus of this work. The initial global two-component decomposition demonstrates that all the spectra can be accurately represented as a linear combination of "cold" and "hot" component sub-spectra (see SI Fig. 1 for further details). The subsequent multi-component analysis is performed by carrying out a two-component SMCR decomposition of pairs of spectra, where for each temperature the "hot" component has a fixed shape (equaling the spectrum at T = 370 K) which can vary in amplitude, while the shape of the "cold" component is allowed to vary with temperature. This modified pairwise approach allows us to see more subtle temperature-dependent variations in the shape of the "cold" component, since it does not assume that all the frequency regions of the "cold" component have identical temperature dependence.

Figure 1c shows the results of the pairwise decomposition, showing the relative amplitudes of the "hot" component (of fixed shape) and the "cold" component of variable shape. The main peak of the "hot" component spectrum is centered at high frequencies associated with weak or broken hydrogen bonds, while the "cold" component is shifted to lower frequencies associated with molecules in a more tetrahedral environment with stronger hydrogen bonds whose population increases with decreasing temperature.[8] However, in addition to the main peak at frequencies around 3200 cm$^{-1}$, the "cold" component displays an additional subtle feature at high frequency, near 3650 cm$^{-1}$. The position of this feature is surprising, since it is located in a region that is usually assigned to molecules with free OH groups i.e. those that do not participate in hydrogen bonds.[11] However, since this appears in the "cold" component, such an assignment would suggest an increase in the number of free OH groups with decreasing temperature. This is highly counterintuitive, since one would expect that free OH groups, which result from broken hydrogen bonds, are entropically stabilized and therefore would decrease in population when water is cooled. While this high frequency feature in the "cold" spectrum is observed both in the global (SI Figure 1) and pairwise analysis, the pairwise SMCR analysis reveals (Figure 1d) that the intensity of the high-frequency feature in the "cold" spectrum exhibits a non-monotonic temperature dependence. In other words, the intensity of this feature initially rises when water is cooled from its boiling point, then reaches a maximum at a temperature of ~280 K, after which it decreases as water is supercooled (Figure 1e).

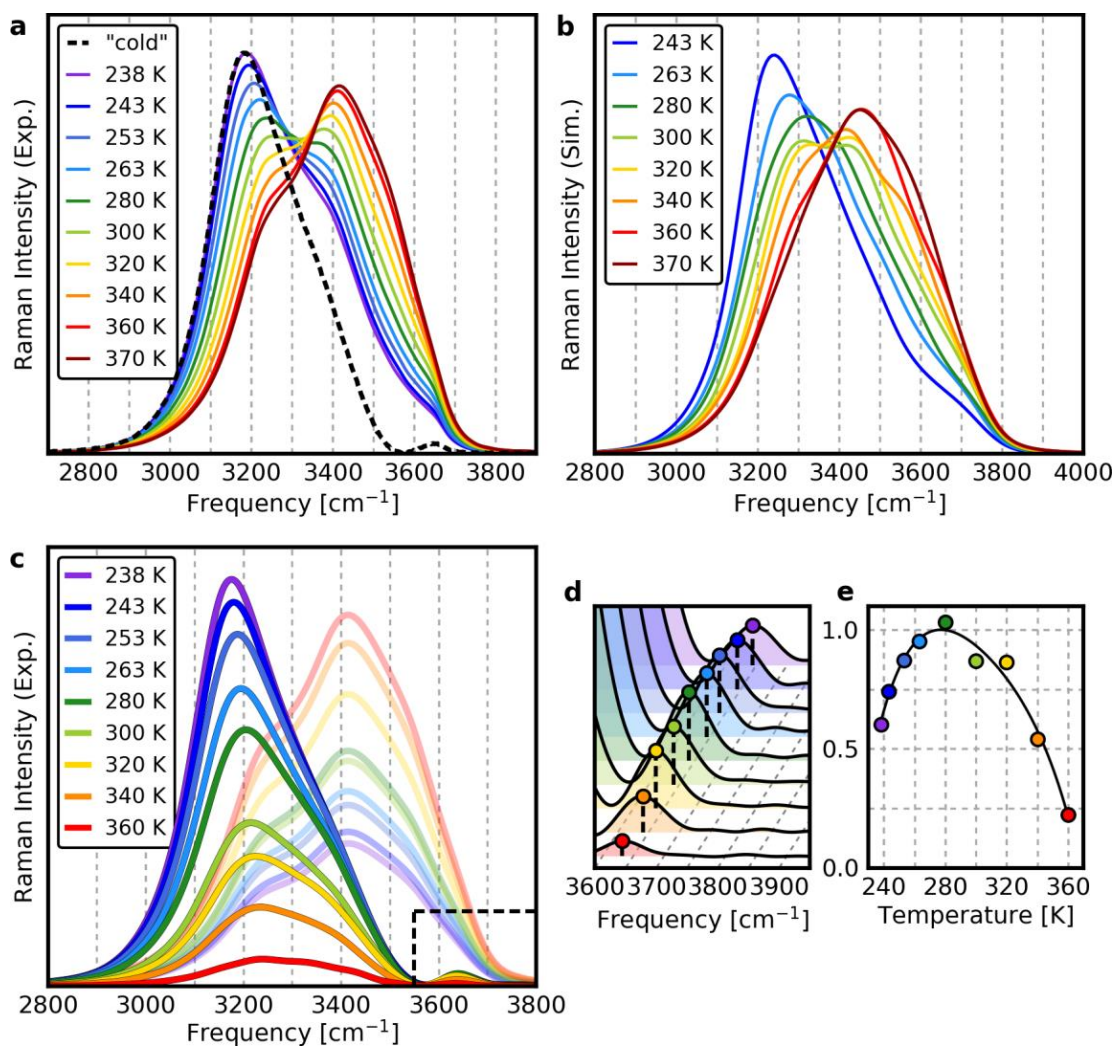

**Figure 1. Analysis of the temperature-dependent O-H stretch band in liquid water. a**, Experimental isotropic Raman spectra showing the bimodal nature of the temperature-dependent O-H stretch band. The "cold" component obtained from a global SMCR decomposition is rescaled to match the maximum of the full spectrum at 238 K and shown as dashed line. **b**, Isotropic Raman spectra obtained from condensed-phase simulations with the machine-learned potential. **c**, Pairwise SMCR decomposition of the experimental Raman spectra into a "hot" component (transparent lines) and a "cold" component (solid lines) revealing a high-frequency feature that lies in a region commonly associated with free O-H stretch vibrations. **d**, Zoom-in into the high-frequency feature where intensity maxima are indicated by filled circles. **e**, Temperature dependence of the intensity maxima, displaying a maximum at a temperature of ~280 K. The solid line is a polynomial fit to the intensity values.

To uncover the molecular origin of the observed spectral features and their temperature dependence, we make use of the additional information provided by the simulations. SI Figure 2a shows that, when pairwise SMCR decomposition is applied to the simulated isotropic Raman spectra, one also finds a distinct high-frequency feature in the "cold" component, which is centered around 3750 cm$^{-1}$ (100 cm$^{-1}$ higher than observed experimentally which is consistent with the shift in the OH region of the spectrum observed in Fig. 1a-b). While the high-frequency feature can be clearly identified, the temperature-dependence of its intensity is statistically noisy, since computing the Raman spectrum from the simulations requires performing computationally expensive electronic structure calculations to obtain the polarizability of the entire system which is only affordable on a subset of structures generated from the simulation. We therefore make use of the vibrational spectrum (vibrational density of states, VDOS). This can easily be calculated for the entire simulation and also allows us to decompose the vibrational spectrum into the local atomic environments that give rise to distinct spectral features. As seen in SI Figure 2b, an SMCR analysis of the calculated vibrational spectrum reveals a high-frequency shoulder on the OH band with a very similar non-monotonic temperature dependence to that of the experimental high-frequency peak, with a maximum at 287 K (close to the experimental maximum at 280K).

Using the simulations we can decompose the vibrational spectrum into the spectra associated with different hydrogen bond environments and use these to establish the origin of the high frequency feature in the "cold" component of the spectrum. We label these environments by the number of hydrogen bonds a molecule donates (D*n*) and accepts (A*m*), respectively[11,21–23] e.g. a molecule in a D2A1 environment donates two hydrogen bonds to neighboring molecules and accepts a single hydrogen bond. Figure 2a shows that when the OH stretch band at 300 K is decomposed into these hydrogen bond environments significant spectral intensities at the high frequency side of the OH stretch only occur for three species: "free" OH's belonging to molecules in D1A1 or D1A2 configurations and the overcoordinated (D2A3) acceptors. However, Figure 2b shows that as the temperature is lowered the former two of these species show a monotonic decrease with temperature while the latter exhibits a maximum which coincides with the temperature dependence of the high-frequency feature observed in the SMCR component of the experimental isotropic Raman spectra. As expected, the component that increases as the temperature is lowered is that associated with water molecules in tetrahedral (D2A2) environments, which gives rise to the main peak of the "cold" spectrum. The spectral shape of the overcoordinated acceptors, together with the temperature dependence of the D2A3 population, provides strong evidence that the high-frequency feature in the "cold" component arises from hydrogen atoms associated with the overcoordinated species.

The simulated vibrational spectrum can be used to predict the other regions that give rise to spectral signatures of overcoordinated hydrogen bonds. To achieve this we use the simulated two-dimensional correlation spectrum[8,24] to assess the vibrations that are strongly correlated with the high-frequency overcoordinated water OH stretch. SI Figure 3, shows that there is a strong positive correlation between the high-frequency overcoordinated region with the low-frequency side of the librational band at ~350 cm$^{-1}$ along with a small positive correlation on the low frequency side of the water bending mode. Performing a SMCR decomposition of the low-frequency part of the vibrational spectrum (Figure 3a) and analyzing the temperature dependence of the intensity at 350 cm$^{-1}$ (Figure 3b-c) reveals the same non-monotonic behavior observed in the OH stretch region (3750 cm$^{-1}$) with a maximum at 294 K. The accuracy of this simulation prediction can be confirmed by SMCR analysis of our experimental anisotropic Raman spectra (SI Figure 4) that shows the onset of an intensity turnover with respect to temperature at frequencies around 350 cm$^{-1}$, albeit less prominently than that in the OH stretch region of the isotropic Raman spectrum due to the high density of other vibrational modes at that frequency.

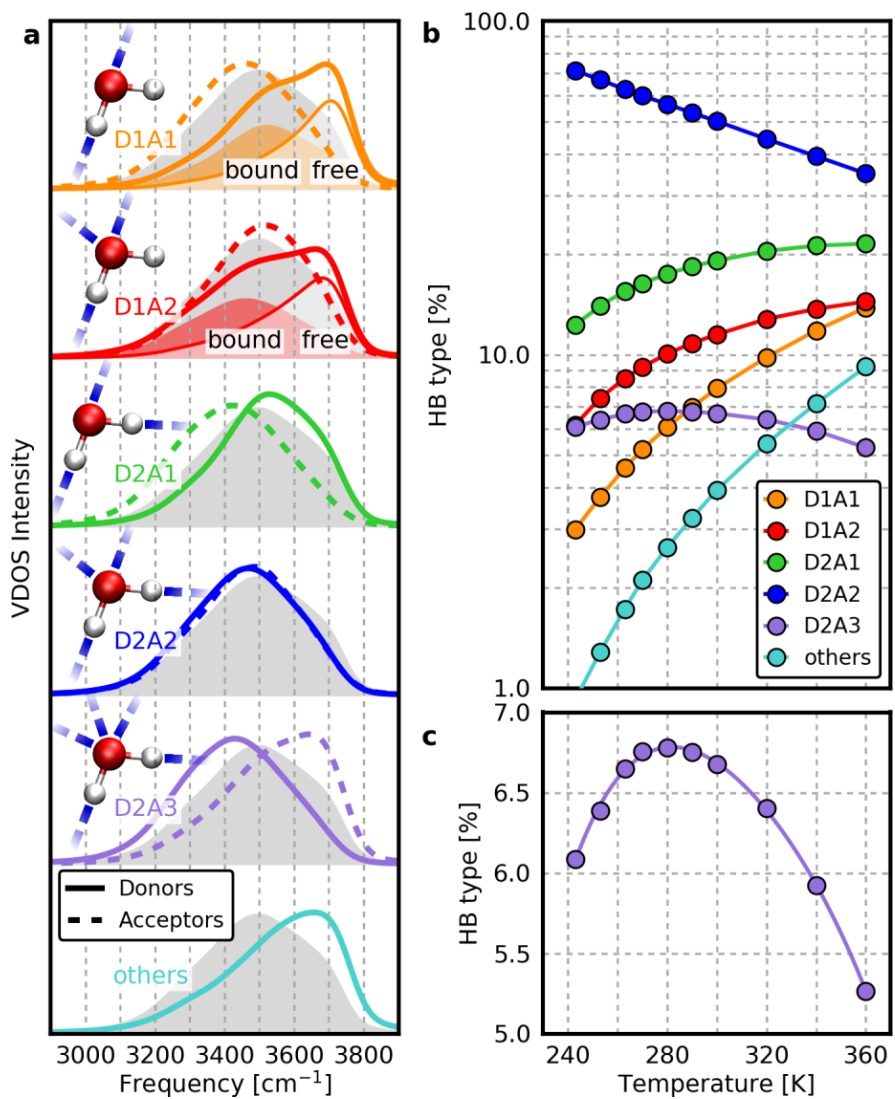

**Figure 2. Decomposition of the simulated O-H stretch band into contributions from donor-acceptor species. a**, Contributions of molecules in different donor-acceptor configurations to the O-H stretch band of the VDOS spectrum at 300 K based on condensed phase simulations with ab initio-based machine learning potentials, showing increased intensity at the blue-side arising from D2A3 acceptor hydrogen bonds. Vibrational spectra of species with a single donating hydrogen are further decomposed into free and bound hydrogens. The full VDOS is shown as grey shading. **b**, Temperature dependence of the different donor-acceptor populations (plotted on a log-scale), showing an initial population increase upon cooling for molecules in tetrahedral (D2A2) and overcoordinated (D2A3) configurations. **c**, Zoom-in into the D2A3 species, revealing a population maximum at a temperature of ~280 K.

To further understand the overcoordinated water molecules we characterized their surrounding hydration environments. The three acceptor hydrogen bonds donating to the overcoordinated (D2A3) water molecules involve primarily species that are in tetrahedral (D2A2) and D2A1 configurations (see SI Fig. 5). However, D2A1 environments are more likely to donate a hydrogen bond to the overcoordinated water molecules relative to their statistical occurrence by a factor of ~1.5 across the entire temperature range. For example at 300 K D2A1 comprises ~30% of the hydrogen bonds accepted by overcoordinated water molecules but at this temperature only ~20% of all the water molecules are in D2A1 environments. While it could appear that overcoordinated molecules arise from short-lived fluctuations in the hydrogen bond network, since a D2A3 and a D2A1 molecule have in total 4 donors and 4 acceptors and therefore a balanced hydrogen bond configuration, the conversion of a D2A3-D2A1 pair into a D2A2-D2A2 structure would require large re-arrangements of the hydrogen bond network (see Fig. 4b). Also, as shown in the SI (SI Fig. 6 and SI Table 1), the lifetimes of the overcoordinated species are 157 fs at 300 K which is comparable to the lifetimes of other species with broken hydrogen bonds and only slightly shorter than the ones observed for tetrahedral environments (245 fs). At 243 K these hydrogen bond environment lifetimes increase to 1341 fs and 2048 fs, respectively.

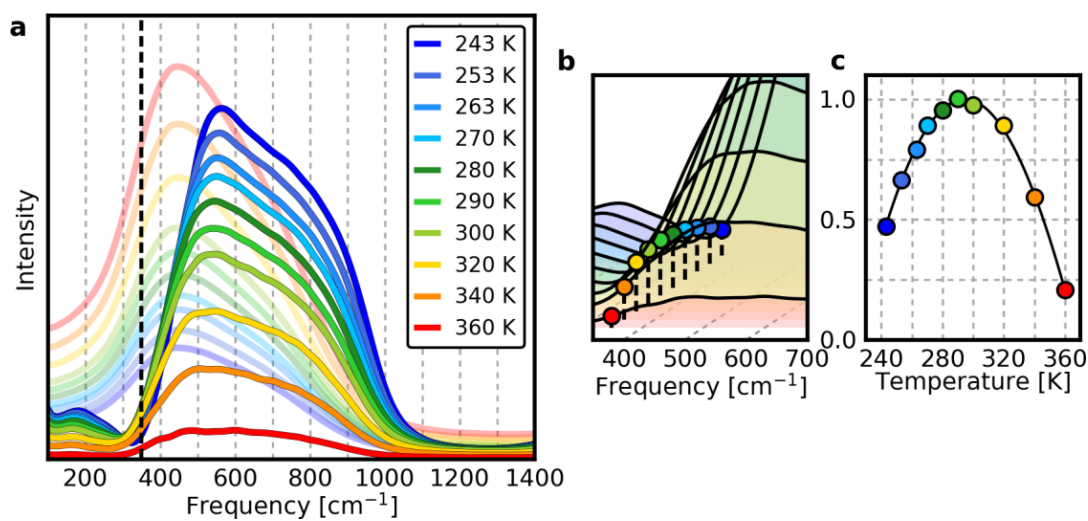

**Figure 3**. **Analysis of the simulated vibrational spectra at low frequencies.** Pairwise SMCR decomposition of the simulated vibrational spectrum in the low-frequency regime into a "hot" component (transparent lines) and a "cold" component (solid lines) **(a,b)**. A cut at a frequency of 350 cm$^{-1}$ (informed by the 2D frequency analysis shown in SI. Figure 3) shows the same temperature dependence as the one found for the high-frequency feature **(c)**.

The geometric properties of the hydrogen bonds formed by overcoordinated molecules also deviate significantly from those in tetrahedral (D2A2) environments. In particular, while the donor and acceptor hydrogen bonds of tetrahedral species possess nearly identical properties, the intramolecular OH bond length show an increase by 5 mÅ for overcoordinated donors and a decrease of 4 mÅ for overcoordinated acceptors relative to the average OH group in water (Figure 4). These bond length shifts are accompanied by donor OH stretches that are red shifted by 57 cm$^{-1}$, in contrast to the blue shift of 56 cm$^{-1}$ of the acceptor hydrogens consistent with the established negative linear correlation between the OH frequency and bond length[25] which has recently been further elucidated.[26] The overcoordinated donor hydrogen bonds also have shorter than average OH…O oxygen-oxygen distances ($r_{OO}$) and hydrogen bond angles ($\theta_{OH…O}$), indicating the formation of stronger hydrogen bonds (Fig. 4c) while the acceptor hydrogen bonds exhibit a considerably wider range of distances and angles (Fig. 4a). However, even though the three acceptor hydrogen bonds have a more diffuse distribution of distances and angles, characteristic of weaker hydrogen bonds, they still satisfy the standard geometric criteria used to identify hydrogen bonds.[27,28] The breadths of the angular distributions observed for the donor and acceptor hydrogen bonds are also consistent with their respective blue and red shifts of the libration band (SI Fig. 3) since this band arises from hindered rotations of the water molecules and is thus correlated with the angular freedom of the hydrogen. Hence despite the presence of three acceptors hydrogens around the overcoordinated water molecule they possess comparatively high rotational freedom. The configuration in which the five neighbors are spatially oriented relative to the central molecule can be described as distorted trigonal bipyramid, as inferred from the bimodal OOO angular distribution between the central and the acceptor oxygens which shows two short angles close to 90° and one large angle close to 180° (see SI Fig. 7).

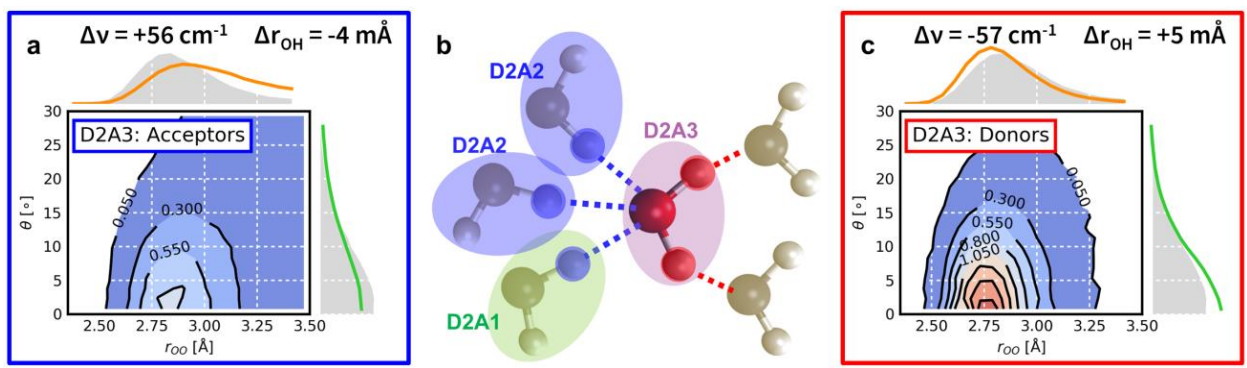

**Figure 4. Characterization of the overcoordinated solvation environment at T = 300 K.** OH groups involved in acceptor hydrogen bonds display blue-shifted frequencies and shortened bond length compared to the average **(a)** and belong primarily to molecules in tetrahedral (D2A2) or D2A1 configurations **(b)**, while donor OH groups display red-shifted frequencies and lengthened bond length **(c)**. $\Delta \nu$ is defined as the difference between the average D2A3 acceptor/donor frequency and the average frequency of the full OH stretch band of the vibrational spectrum. $\Delta r_{OH}$ is defined as the difference between the average D2A3 acceptor/donor intramolecular OH bond length and the OH bond length averaged over all molecules. **a**, Probability density function of the hydrogen bond angle θ and the oxygen-oxygen distance $r_{OO}$ of the D2A3 acceptor bonds, showing hydrogen bonds of weak, delocalized nature. **c**, Analogous analysis for D2A3 donor bonds, displaying a more localized distribution with shorter distances and angles, indicating the formation of strong hydrogen bonds. For reference the grey shaded distributions in **a** and **c** are averaged over all hydrogen bonds.

Since overcoordinated environments exhibit a high local density and their temperature of maximum population is close to the temperature of maximum density of liquid water, one might assume that the temperature dependence of these environments simply follows the density variation of the liquid with temperature. To test this hypothesis, we performed simulations over the same temperature range (T = 240 K to 370 K) while holding the density constant (at ρ = 0.997 g/cm$^3$). As shown in SI Fig. 8, the maximum in the overcoordinated water population still occurs under these constant-density conditions and its temperature dependence (along with the temperature dependence of the other hydrogen bond species) is essentially unchanged even when the total density of the system is held fixed. These results clearly demonstrate that overcoordinated environments in liquid water are not merely a consequence of the water density but conversely is likely an indicator of the underlying process that causes the change in water density with temperature, as the non-monotonic overcoordinated (D2A3) population, with high local density, would be expected to favor higher total densities under constant pressure conditions. In other words, the maximum in the overcoordinated population lead to pressure minimum in the constant-density simulation, and that pressure minimum becomes a density maximum when the system is held at constant pressure. Our results are also consistent with prior studies suggesting[20,29] that the density maximum is caused by the balance between the first coordination shell moving closer and the second hydration shell moving away from the central molecule upon cooling. This leads to a drop in the number of interstitial molecules that penetrate the first solvation shell and hence reduces the formation of overcoordinated environments. Indeed, as shown in SI Fig. 9 while the four nearest neighbor oxygens of the overcoordinated water molecules are at similar distances to those found in tetrahedral configurations and exhibit similar temperature dependence, their fifth neighbors are shifted to much closer distances and exhibit the opposite temperature dependence to the other hydrogen bond environments, moving in as temperature is lowered.

In summary, by performing Raman measurements and simulations of water both in the ambient and supercooled regimes we find that features in the OH stretch and librational bands provide spectroscopic signatures that can be used to identify water molecules in overcoordinated hydrogen bonded environments, which donate two strong hydrogen bonds and accept three weak hydrogen bonds. Despite satisfying the geometrical criteria usually associated with hydrogen bond formation, the spectroscopic signatures of the overcoordinated acceptor OH groups are found to lie in regions more commonly associated with non-hydrogen bonded OH groups, making it easy to misinterpret spectra as probing undercoordinated environments rather than the opposite. Given the role of overcoordinated environments in many important aqueous processes these findings offer the experimental and theoretical means to identify and interpret such structures in chemical and biological systems.

**Acknowledgements**

This material is based upon work supported by the National Science Foundation under Grant No. CHE-1652960 to T.E.M. and Grant No. CHE-1464904 to D.B.A. T.E.M. also acknowledges support from the Camille Dreyfus Teacher-Scholar Awards Program. T.M. is grateful for financial support by the DFG (MO 3177/1-1). We thank Ondrej Marsalek, Andrés Montoya-Castillo and Yuezhi Mao for useful discussions.

SUPPORTING INFORMATION FOR:

# Hiding in the crowd: Spectral signatures of overcoordinated hydrogen bond environments

Tobias Morawietz[a], Andres S. Urbina[b], Patrick K. Wise[b], Xiangen Wu[c], Wanjun Lu[c,] Dor Ben-Amotz[b], and Thomas E. Markland[a,*]

a) Department of Chemistry, Stanford University, Stanford, CA 94305, United States

b) Department of Chemistry, Purdue University, West Lafayette, IN 47907, United States

c) College of Marine Science and Technology, China University of Geosciences, Wuhan 430074, China

*Email: tmarkland@stanford.edu

## 1. Experimental details

Polarized Raman spectra of water at ambient pressure and temperatures down to T = 0 C°, were obtained using a 514.5 nm Ar-ion excitation laser with ~20 mW at the sample and 5 minute integration time per spectra, as previously described.[1] Polarized spectra at supercooled temperatures were obtained using 532.06 nm doubled Nd:YAG laser with ~10 mW at the sample, contained in a Chou-Burruss-Lu type optical capillary cell[2] as previously described.[3,4] The only new feature of the latter instrument is that polarizers were placed in the input laser path and Raman detection paths in order to collect the parallel and perpendicularly polarized Raman spectra from which the isotropic Raman spectra were obtained as previously described.[1]

## 2. Simulation details

The machine-learned potential used in this work is based on artificial neural network potentials employing the Behler-Parrinello approach[5–9] and was constructed and validated in Ref.[9] The potential was trained to density-functional theory calculations for a set of 9,189 condensed-phase liquid water configurations using the revPBE exchange-correlation functional[10,11] with D3 dispersion correction[12] employing specific settings as previously reported.[9,13] Simulated Raman spectra were obtained from autocorrelation functions of polarizability tensor elements as described in Ref.[9] The polarizability tensor of the system was calculated from density functional perturbation theory[14] employing the CP2K code.[15,16] Molecular dynamics (MD) simulations with the machine-learned potential were performed with a modified version of the LAMMPS[17,18] package, using simulation cells containing 128 water molecules, for 2 ns per temperature at T < 300 K and 1 ns otherwise. The standard hydrogen bond criterion[19,20] was used to define the hydrogen bond species. Vibrational spectra associated with different hydrogen bond environments were calculated employing the time-dependent vibrational density of states as described in Refs.[9,13,21]

## 3. Comparison of water models

To confirm the robustness of our identification of the high frequency feature as arising from overcoordinated (D2A3) water molecules we performed additional simulations using the *ab initio*-based MB-pol[22–24] and the rigid, empirical TIP4P/2005[25] models. These simulations display hydrogen-bond populations that are in close agreement with the ones obtained with the machine-learned potential (see SI Fig. 10) and also show qualitatively the same temperature dependence for the overcoordinated species, with population maxima at 280 K and 296 K, respectively. Similarly a vibrational spectrum decomposition of the MB-pol spectrum and an electric field distribution analysis of the TIP4P/2005 simulations (see SI Figure 11) both confirm our findings: donor and acceptor spectra of molecules in tetrahedral environments are indistinguishable while acceptor hydrogens of overcoordinated species give rise to spectral signatures that are shifted to high frequencies (low electric field strengths). MD Simulations with the MB-pol[22–24] model were carried out for 200 ps per temperature employing the i-PI code[26] with simulation cells of 256 water molecules. MD Simulations with the TIP4P/2005[25] model were performed employing the simulation package GROMACS[27,28] for 1 ns per temperature using simulation cells containing 512 water molecules.

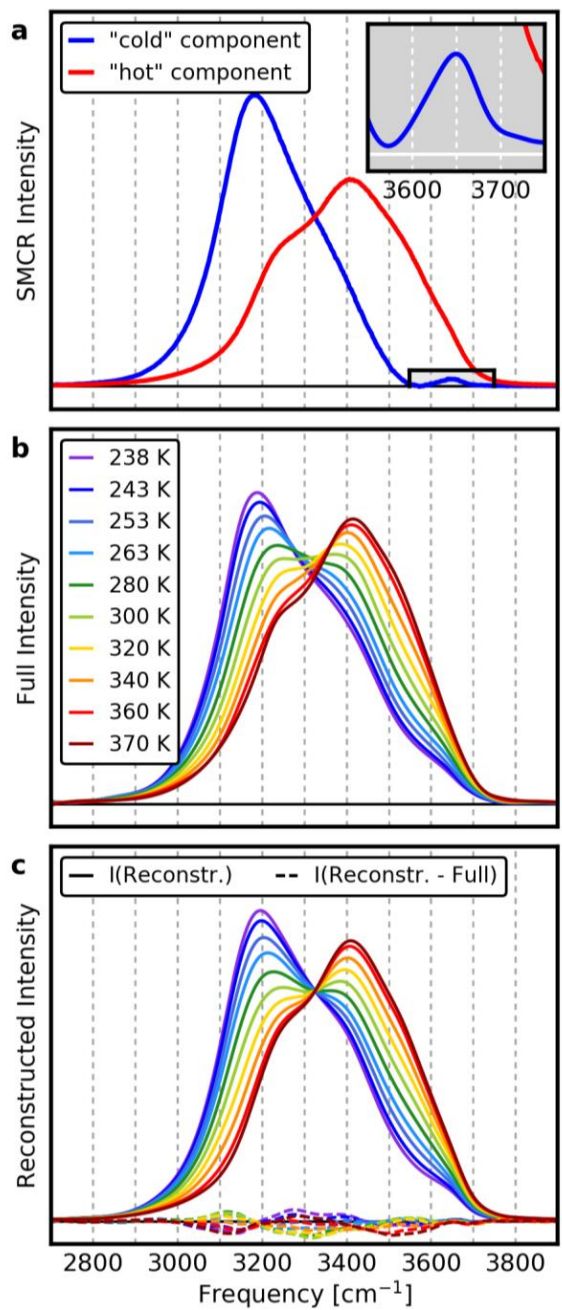

**SI Figure 1. Analysis of the experimental Raman O-H stretch band. a**, Global SMCR decomposition of the temperature-dependent isotropic Raman spectra into a "cold" and a "hot" component whose shapes are both independent of temperature. The inset shows the high-frequency feature of the "cold" component whose position coincides with the free O-H stretch band. **b**, Measured isotropic Raman spectrum as function of temperature. **c**, Reconstructed spectrum obtained by combining the temperature-independent SMCR components multiplied by their population. The intensity difference with respect to the full spectrum is indicated by dashed lines.
4

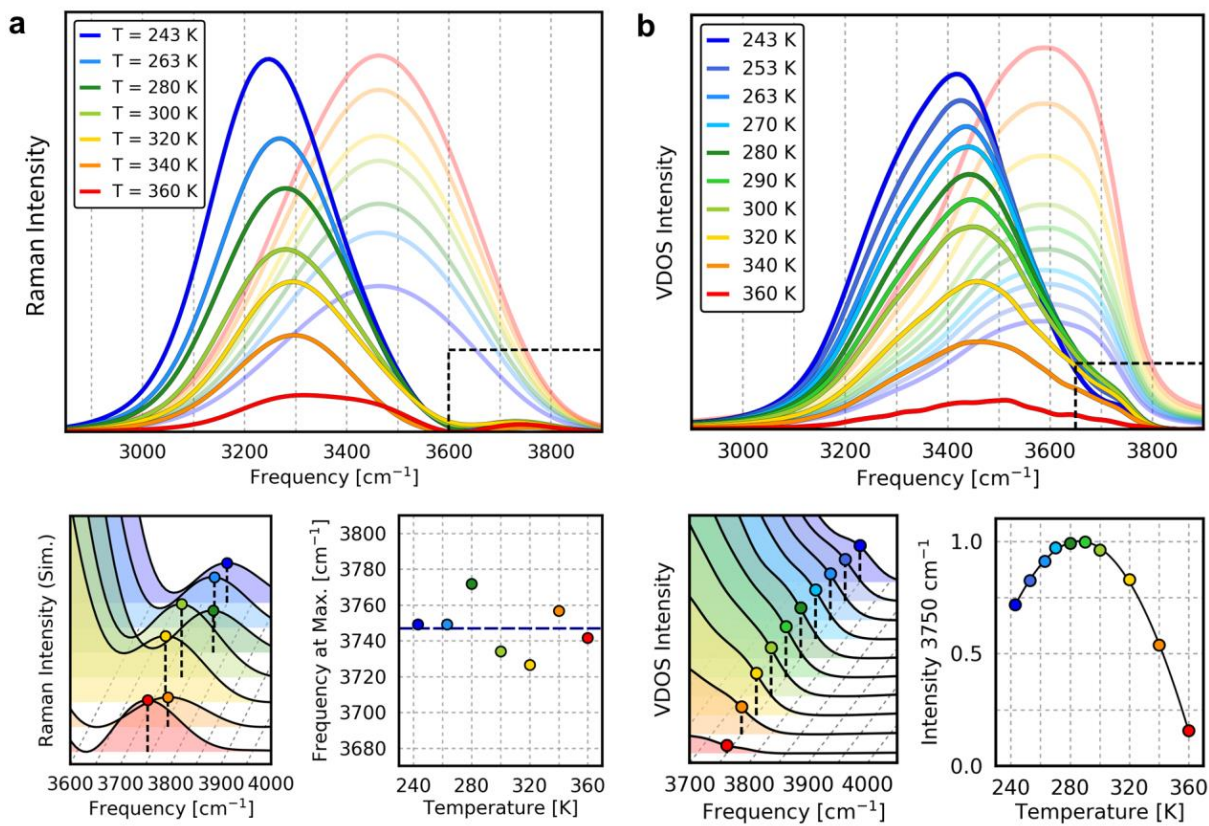

**SI Figure 2. Analysis of the simulated Raman and simulated VDOS O-H stretch band. a**, Pairwise SMCR decomposition of the simulated isotropic Raman spectra into a "hot" component (transparent lines) and a "cold" component (solid lines), revealing a high-frequency peak centered at 3750 cm$^{-1}$. **b**, Equivalent analysis for the simulated VDOS spectra, revealing a high-frequency shoulder. A cut through this shoulder at a frequency of 3750 cm$^{-1}$ shows a non-monotonic temperature-dependence of the intensity with a maximum at ~290 K.

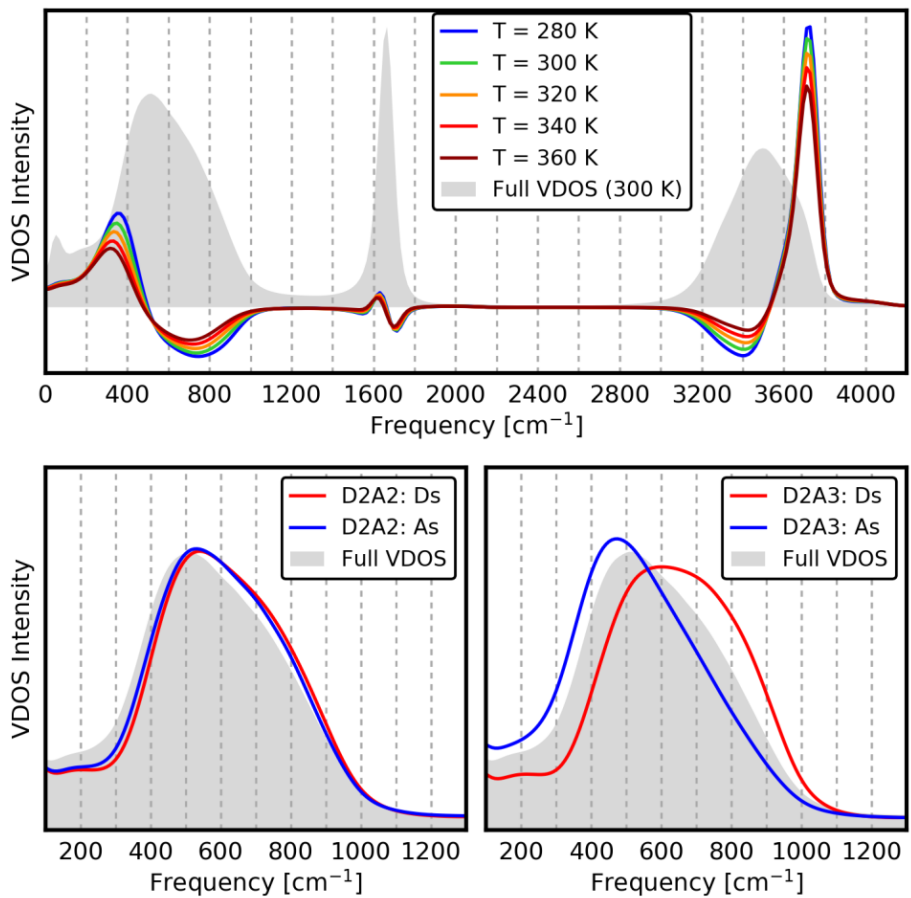

**SI Figure 3. Analysis of the low-frequency regime. (Top),** Cut through the synchronous two-dimensional correlation spectrum of the simulated VDOS at frequencies of 3750 cm$^{-1}$ reveals positive correlation in the librational regime at ~350 cm$^{-1}$. **(Bottom)** Decomposition of the VDOS at low-frequencies shows that donor and acceptors Hydrogens belonging to D2A2 molecules give rise to indistinguishable spectral features while for the D2A3 species the acceptor spectrum is shifted to lower frequencies and the donor spectrum is shifted to higher frequencies.

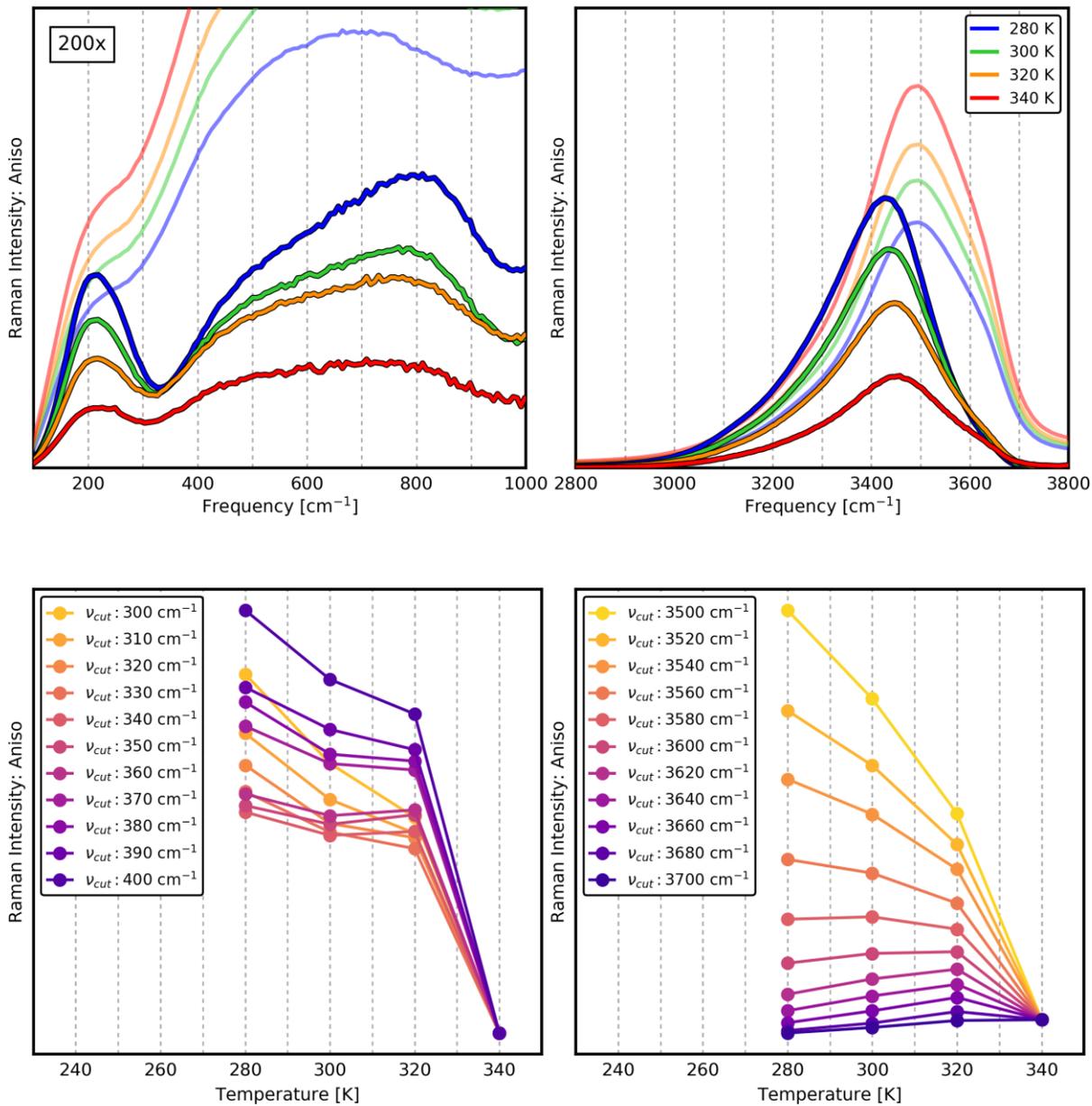

**SI Figure 4. SMCR analysis of the anisotropic Raman spectrum.** Pairwise SMCR decomposition of the experimental anisotropic Raman spectrum in the low-frequency **(top left)** and high-frequency regimes **(top right)**. Variation of the Raman intensity with temperature at fixed frequency values between 300 cm$^{-1}$ and 400 cm$^{-1}$ **(bottom left)**, and between 3500 cm$^{-1}$ and 3700 cm$^{-1}$ **(bottom right)**.

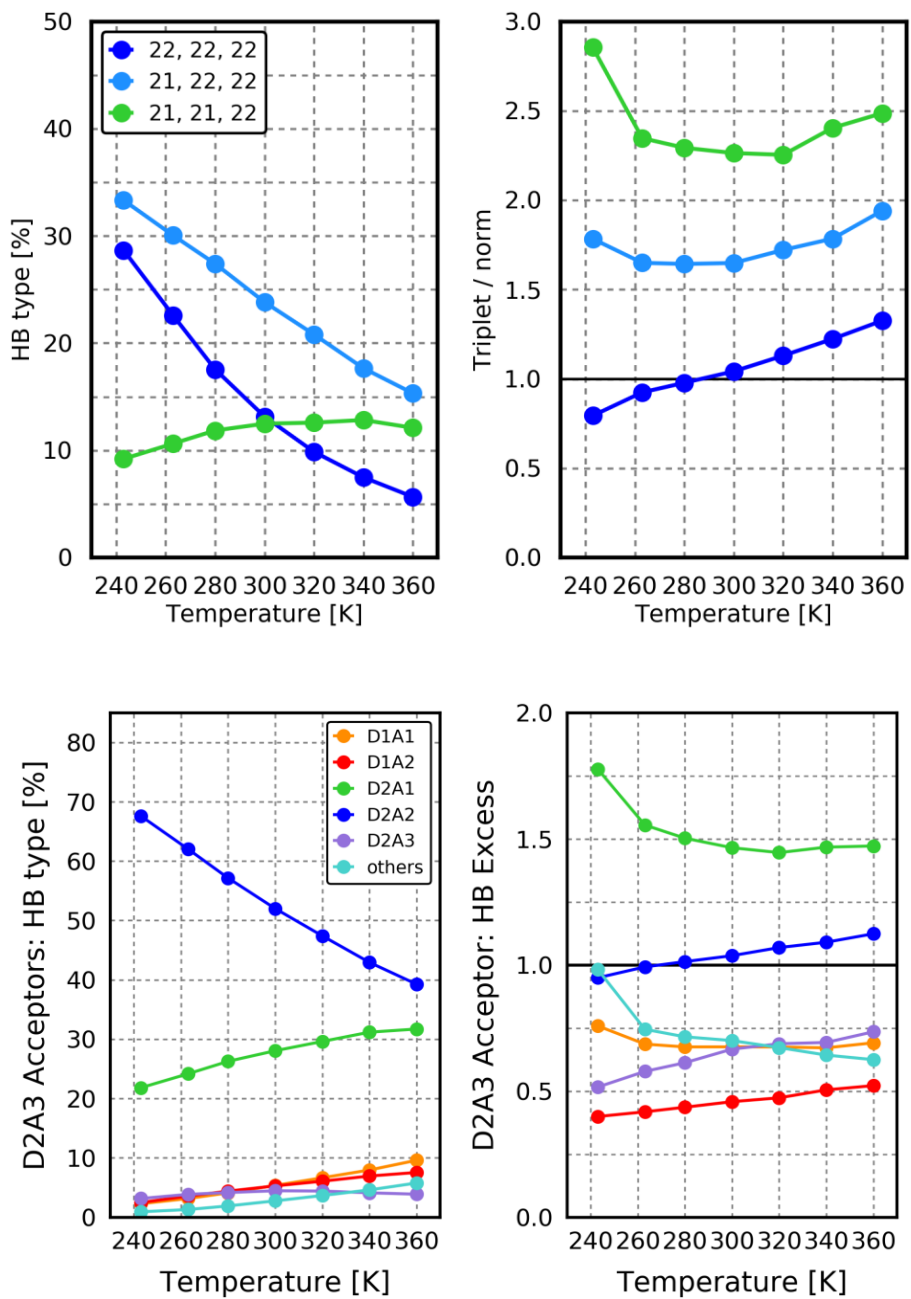

**SI Figure 5. Analysis of acceptor hydrogen bonds donating to overcoordinated species. (Top left)** The three acceptor triplet combinations with the highest probability over the whole temperature range involve hydrogen bond species in D2A2 (22) and D2A1 (21) configurations. **(Top right)** Populations on the top left panel normalized by the number of available species at a given temperature. **(Bottom left)** Temperature dependence of the individual donor-acceptor populations that donate to the overcoordinated (D2A3) species. **(Bottom right)** Populations on the bottom left panel normalized by the number of available species at a given temperature.

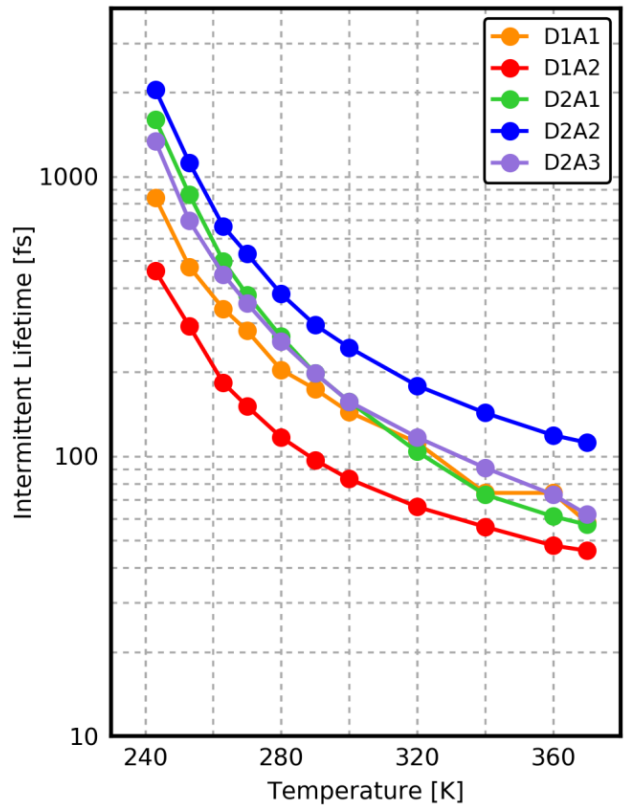

**SI Figure 6. Lifetimes of hydrogen bond species.** Intermittent lifetimes of the different hydrogen bond species obtained from integrating tri-exponential fits to the respective existence autocorrelation functions.

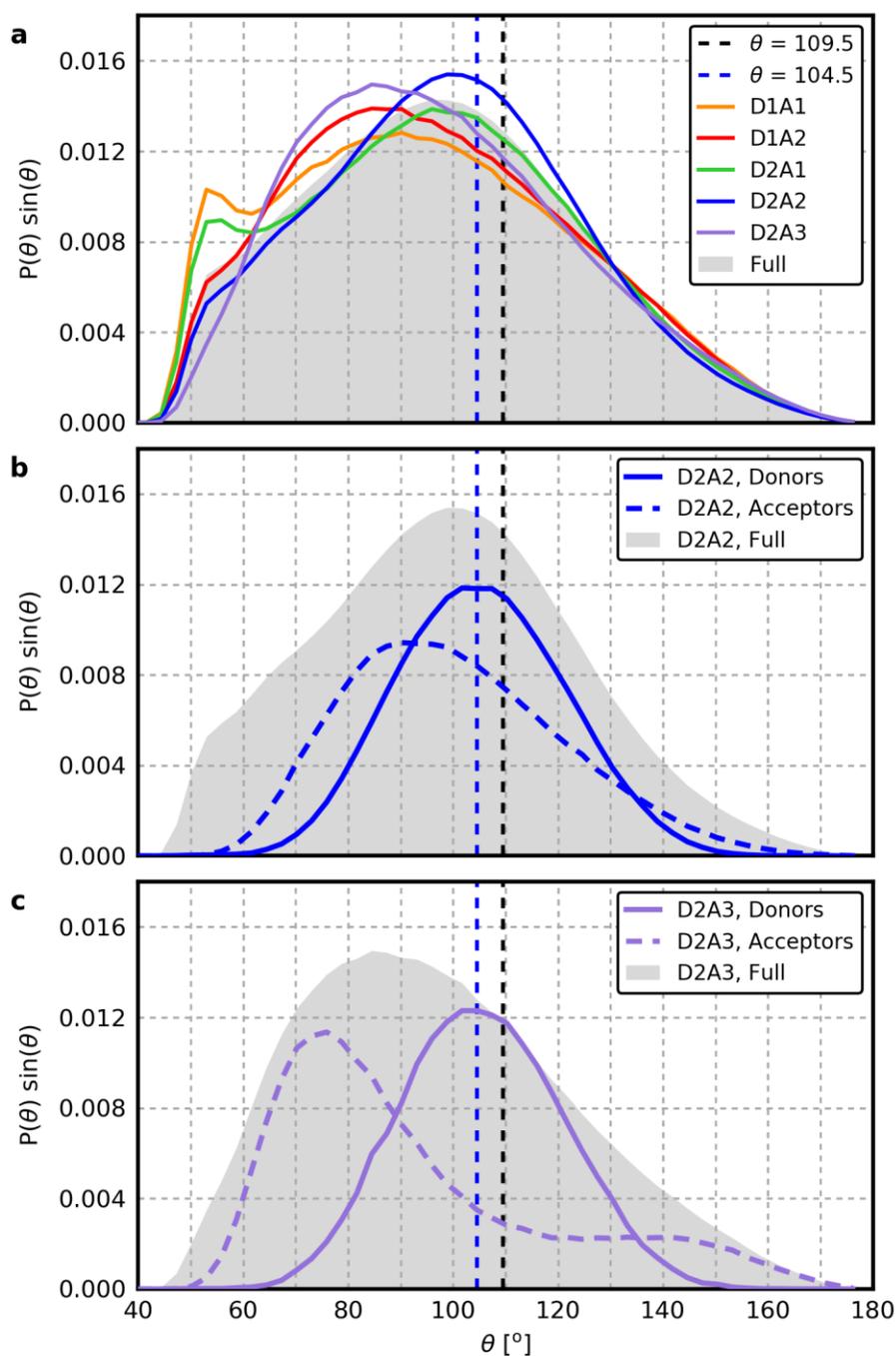

**SI Figure 7. Angle distributions.** Distribution functions of the angle formed between the oxygen of a central water molecules and two of its neighbors. **a**, Distribution functions considering all neighbors within an Oxygen-Oxygen distance cutoff of 3.5 Å for the different hydrogen bond species. **b**, Distributions functions for tetrahedral (D2A2) configurations obtained from considering donor and acceptors oxygens separately. **c**, Distributions functions for overcoordinated (D2A3) configurations obtained from considering donor and acceptors oxygens separately.

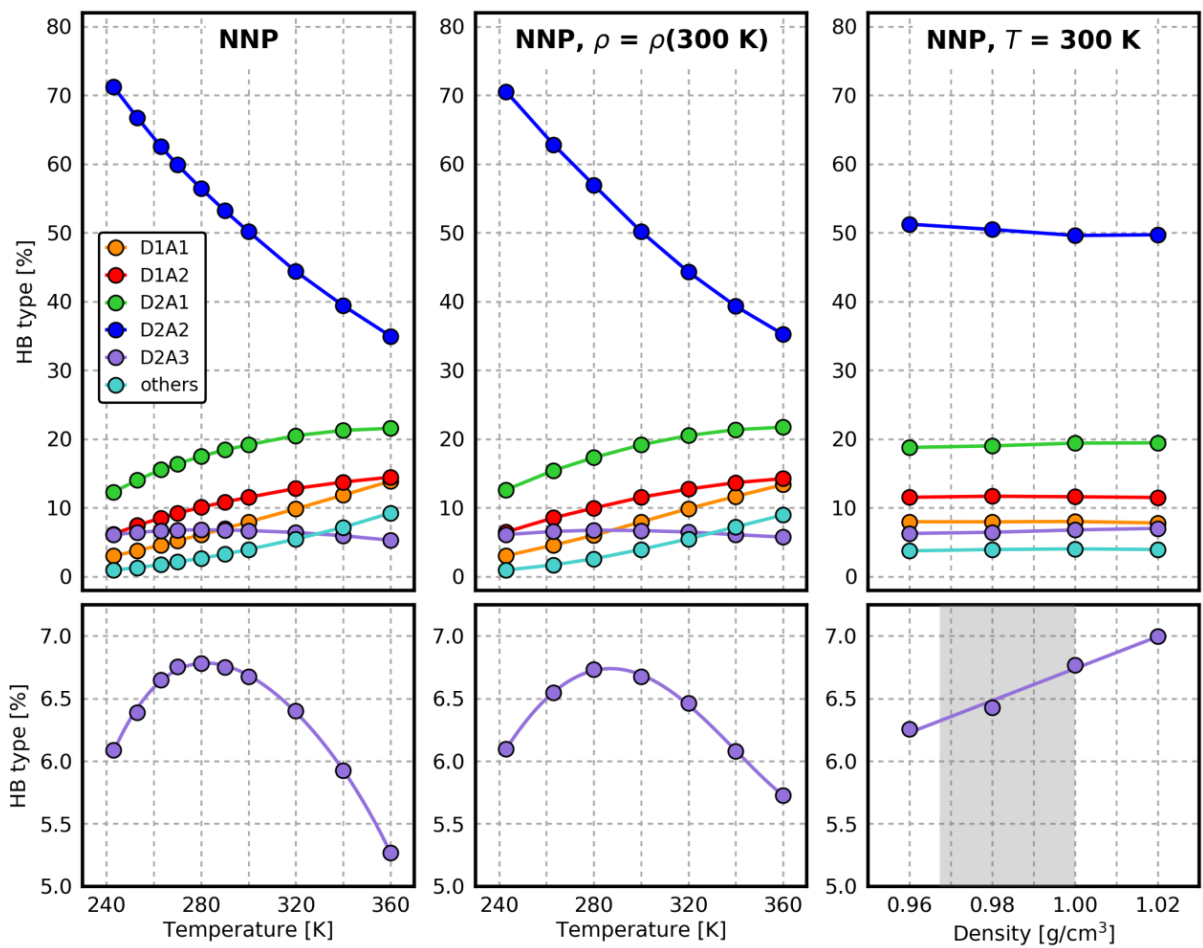

**SI Figure 8. Dependence of hydrogen bond populations on density and temperature. (Left),** Populations obtained from simulations run at the experimental density for each temperature. **(Middle)**, Populations obtained from simulations run at a fixed density for all temperatures ($\rho$ = 0.9966 g/cm$^3$) corresponding to the experimental density at T = 300 K. **(Right)**, Populations obtained from simulations at different densities at a constant temperature of T = 300 K. The shaded area indicates the experimental density range that water occupies at temperatures between 240 K and 360 K.

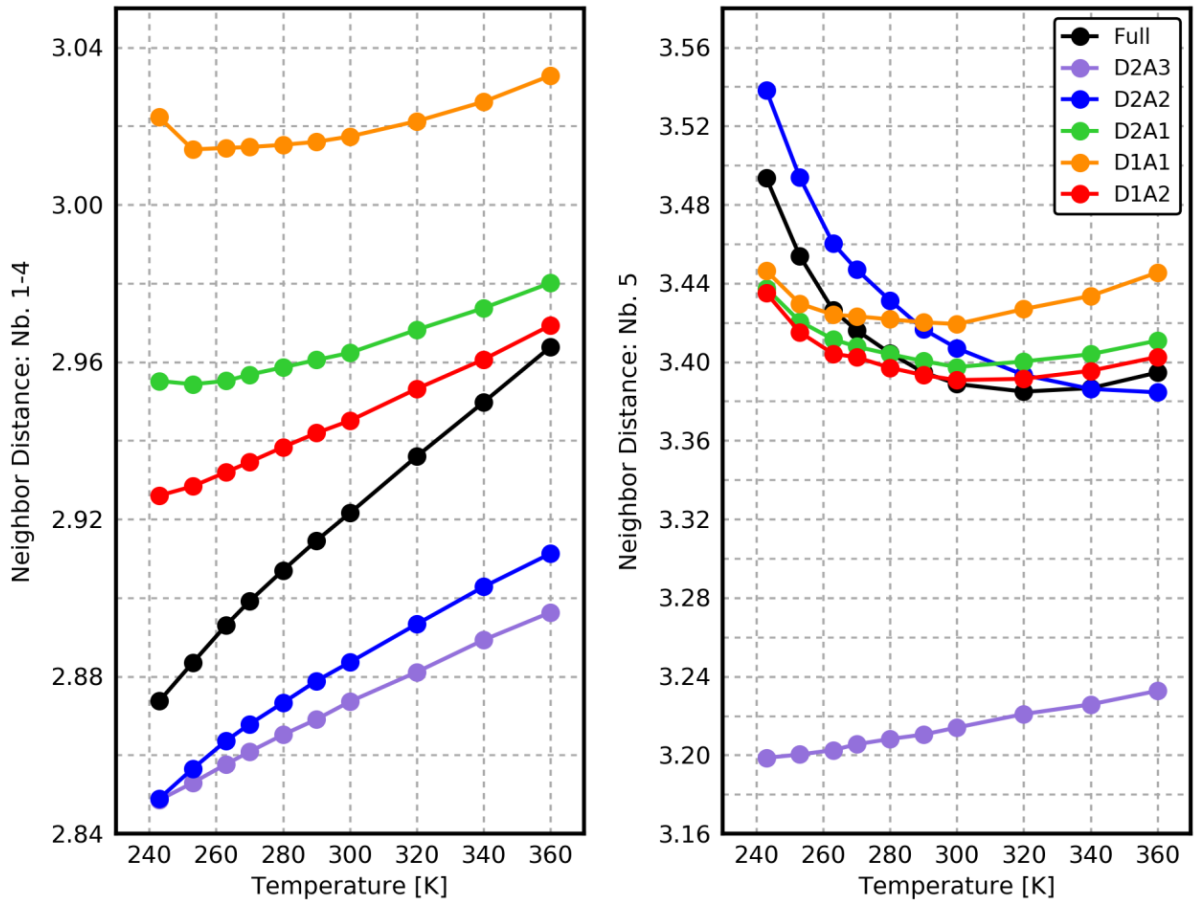

**SI Figure 9. Neighbor distributions.** Average of the Oxygen-Oxygen distribution functions decomposed into contributions from first through fourth nearest neighbor (**left**) and fifth neighbor (**right**) for the different hydrogen bond species as function of temperature.

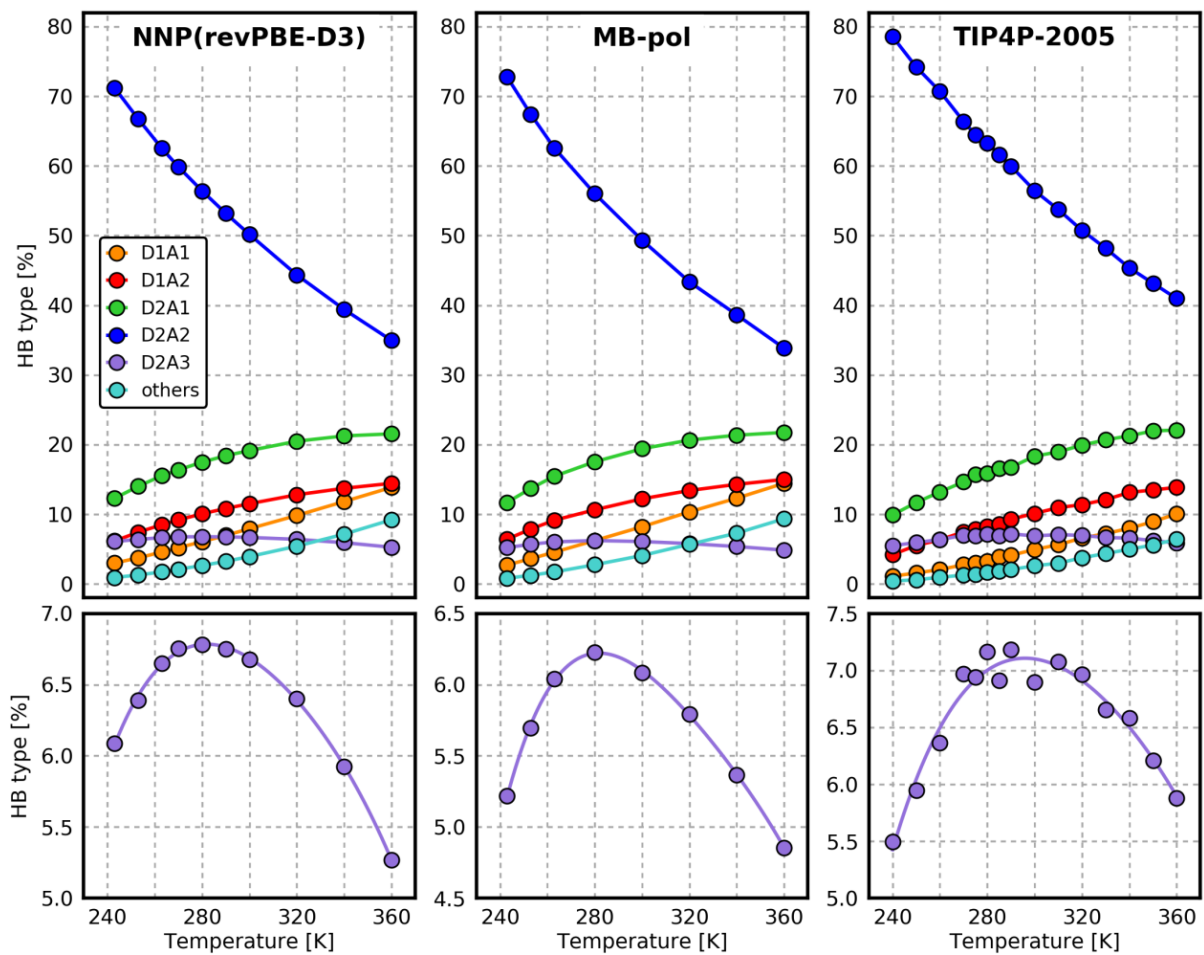

**SI Figure 10. Dependence of hydrogen bond populations on water model. (Top),** Temperature dependence of donor-acceptor species obtained from simulations with the NNP based on revPBE-D3, the ab initio-based MB-pol model, and the empirical TIP4P/2005 force field. **(Bottom)**, Zoom-in into the D2A3 species, revealing a population maximum for all water models.

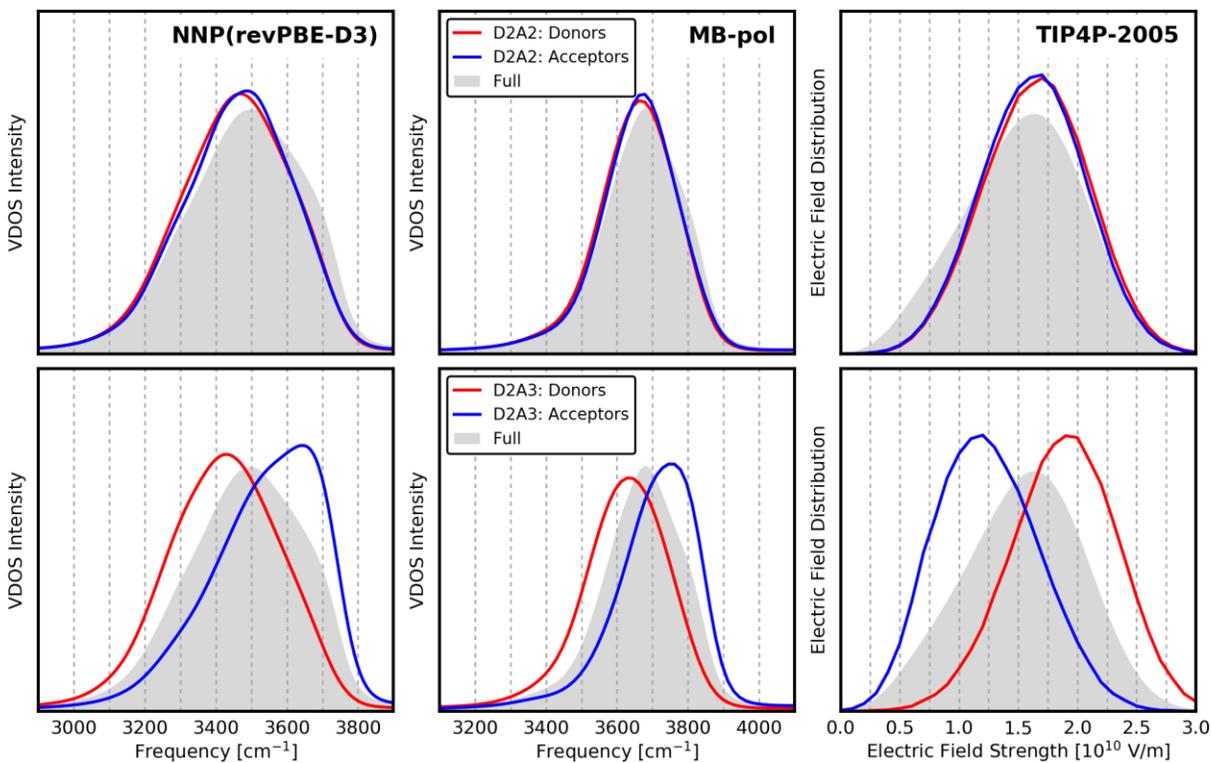

**SI Figure 11. Dependence of spectral decomposition on water model.** Decomposition of the OH stretch band into contributions from D2A2 (top) and D2A3 (bottom) acceptor and donor hydrogens. Decomposition of the VDOS obtained with NNP **(Left)** and MB-pol **(Middle)**, and the electric field distribution obtained with TIP4P/2005 which is approximately linearly related to the OH stretch frequency with high electric field strength corresponding to low frequencies **(Right).**

**SI Table 1.** Numerical values of the lifetimes found at T = 300 K and T = 243 K.

| Species | Lifetime, 300 K [fs] | Lifetime, 243 K [fs] |
|---------|----------------------|----------------------|
| D1A1    | 144                  | 842                  |
| D1A2    | 83                   | 460                  |
| D2A1    | 157                  | 1601                 |
| D2A2    | 245                  | 2048                 |
| D2A3    | 157                  | 1341                 |